\documentclass[12pt,twoside,a4paper]{irs}
\usepackage{epsf,exscale,times}
\usepackage[english]{babel}
\usepackage{graphicx}
\usepackage{amsmath}
\usepackage{fancyhdr}
\usepackage[outdir=./figures/]{epstopdf}

\usepackage{cite}
\usepackage{mathtools,amssymb,amsfonts}
\usepackage{siunitx}
\usepackage{algorithmic}
\usepackage{graphicx}
\usepackage{textcomp}
\usepackage{xcolor}
\usepackage{tabto}
\usepackage{pgfplots}					
\usepgfplotslibrary{statistics}		
\pgfplotsset{width=7cm,compat=newest}	
\newcommand\TP{\mathit{TP}}
\newcommand\FN{\mathit{FN}}
\newcommand\FP{\mathit{FP}}

\usepackage{todonotes}
\def\BibTeX{{\rm B\kern-.05em{\sc i\kern-.025em b}\kern-.08em
    T\kern-.1667em\lower.7ex\hbox{E}\kern-.125emX}}
\usetikzlibrary{external}
\newcommand {\vsp}   {\vspace*}

\def\title#1{\vsp{-16mm}\begin{center}\Large\bf{#1}\end{center}\vsp{0mm}}
\def\author#1{{\begin{center}\textbf{#1}\end{center}\vspace{-1mm}}}
\def\address#1{\vsp{-3mm}\begin{center}\baselineskip12pt\normalsize{#1}\end{center}\vsp{-1mm}}

\def\abstract#1{{\vspace{-5mm}
    \begin{center}
      \begin{minipage}{0.85\textwidth}
        \noindent\bf \textit{Abstract:}
        \small\rm\emph{#1}
				\vsp{-0.5em}
      \end{minipage}
    \end{center}
}}

\def\authorsheadline=#1{\global\def\@authorsheadline{#1}}
\global\def\@authorsheadline{}

\def\TeX{T\kern-.1667em\lower.5ex\hbox{E}\kern-.125emX}

\def\LaTeXG{{\rm L\kern-.36em\raise.3ex\hbox{\sc a}\kern-.15emT\kern-.1667em\lower.7ex\hbox{E}\kern-.125emX}}
\def\LaTeXK{{\it L\kern-.24em\raise.4ex\hbox{\scriptsize \it A}\kern-.20emT\kern-.1667em\lower.5ex\hbox{E}\kern-.125emX}}

\begin{document}

\fancypagestyle{firststyle}
{
   \fancyhf{}
   \rfoot{ \footnotesize {\thepage} }
}

\thispagestyle{firststyle}
\fancyhf{}
\renewcommand{\headrulewidth}{0pt}
\renewcommand{\footrulewidth}{1pt}
\renewcommand{\footskip}{50pt}

\pagestyle{fancy}
\fancyfoot[RO,LE]{ \footnotesize {\thepage} }

\title{Automated Ground Truth Estimation For Automotive Radar Tracking Applications With Portable GNSS And IMU Devices}


\author{
Nicolas Scheiner$^{*}$, Stefan Haag$^{*}$, Nils Appenrodt$^{*}$, Bharanidhar Duraisamy$^{*}$, J\"urgen Dickmann$^{*}$, Martin Fritzsche$^{*}$, Bernhard Sick$^{**}$}
\address{
    $^{*}$Daimler AG\\
    Ulm, Germany\\
		email: \{nicolas.scheiner, stefan.s.haag\}@daimler.com\\[2mm]
    $^{**}$University of Kassel\\
    Kassel, Germany\\
		email: bsick@uni-kassel.de
		}

\abstract{
Baseline generation for tracking applications is a difficult task when working with real world radar data.
Data sparsity usually only allows an indirect way of estimating the original tracks as most objects' centers are not represented in the data.
This article proposes an automated way of acquiring reference trajectories by using a highly accurate hand-held global navigation satellite system (GNSS).
An embedded inertial measurement unit (IMU) is used for estimating orientation and motion behavior.
This article contains two major contributions.
A method for associating radar data to vulnerable road user (VRU) tracks is described.
It is evaluated how accurate the system performs under different GNSS reception conditions and how carrying a reference system alters radar measurements.
Second, the system is used to track pedestrians and cyclists over many measurement cycles in order to generate object centered occupancy grid maps.
The reference system allows to much more precisely generate real world radar data distributions of VRUs than compared to conventional methods.
Hereby, an important step towards radar-based VRU tracking is accomplished.
}


\section{Introduction}
\begin{figure}[tb]
   \centering
   \includegraphics[width=\textwidth]{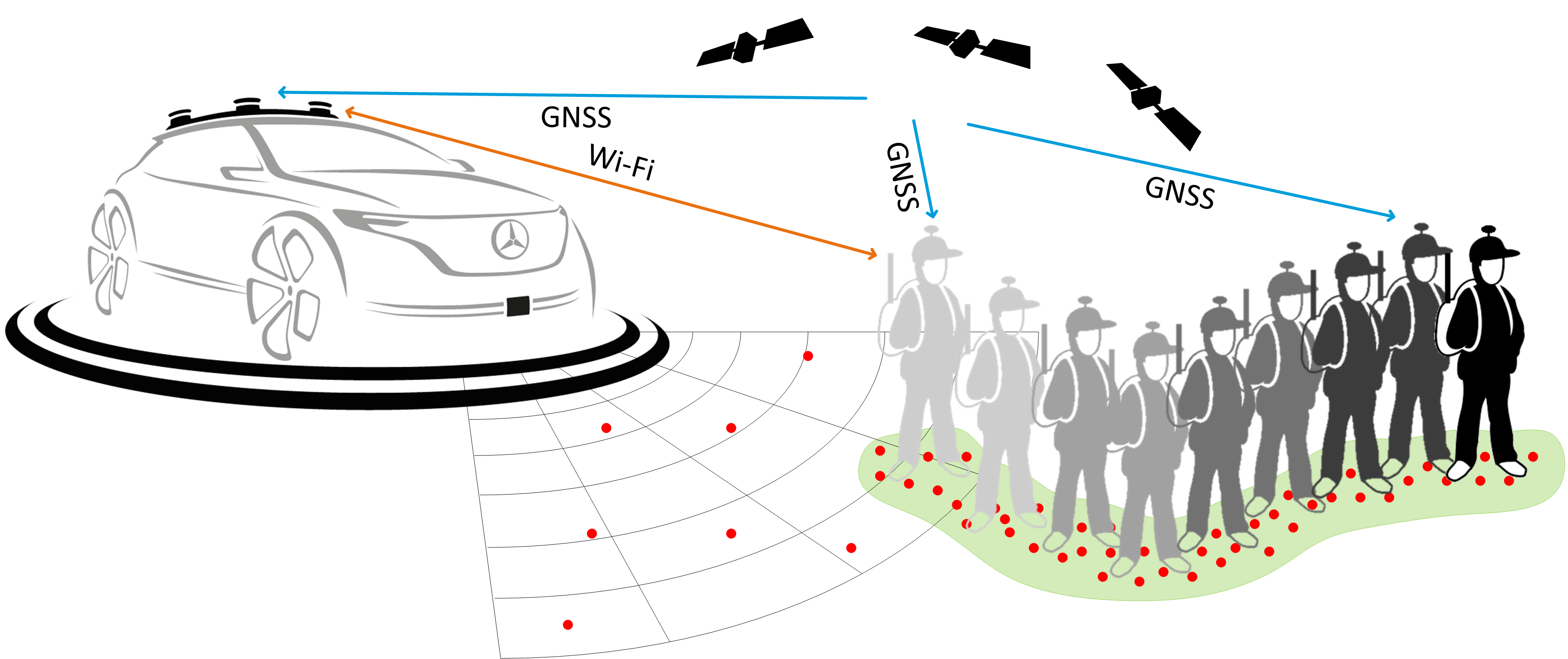}
   \caption{Automated ground truth estimation: GNSS positions of ego-vehicle and VRU are used to automatically determine the object center and an enclosing area in close proximity around  the VRU's location.}
	 \label{fig:sys_overview}
\end{figure} 
Autonomous driving is one of the major topics in current automotive research.
In order to achieve excellent environmental perception various techniques are being investigated.
\emph{Extended object tracking} (EOT) aims to estimate length, width and orientation in addition to position and state of movement of other traffic participants and is, therefore, an important example of these methods.
In the radar domain, research is usually focused on using data on a detection level, such as in \cite{Granstrom2017} or \cite{Haag2018}.
Major problems of applying EOT to radar data are a higher sensor noise, clutter and a reduced resolution compared to other sensor types. 
Among other issues, this leads to a missing ground truth of the object's extent when working with non-simulated data.
A workaround could be to test an algorithm's performance by comparing the points merged in a track with the data annotations gathered from data labeling.
The data itself, however, suffers from occlusions and other effects which usually limit the major part of radar detections to the objects edges that face the observing sensor.
The object center can either be neglected in the evaluation process or it can be determined manually during the data annotation, i.e., labeling process.
For abstract data representations as in this task, labeling is particularly tedious and expensive, even for experts.
As estimating the object centers for all data clusters introduces even more complexity to an already challenging task, alternative approaches for data annotation become more appealing.
To this end, this article proposes using a hand-held highly accurate global navigation satellite system (GNSS) which is referenced to another GNSS module mounted on a vehicle (cf. Fig. \ref{fig:sys_overview}).
The portable system is incorporated in a backpack that allows being carried by vulnerable road users (VRU) such as pedestrians and cyclists.
The GNSS positioning is accompanied by an inertial measurement unit (IMU) for orientation and motion estimation.
This makes it possible to determine relative positioning of vehicle and observed object and, therefore, associate radar data and corresponding VRU tracks.
In \cite{Scheiner2019}, the same hardware setup is used for automated data labeling of individual radar points in order to create a data set for machine learning applications. 
It was found that the internal position estimation filter which fuses GNSS and IMU is not well equipped for processing unsteady VRU movements, thus only GNSS was used there.
In comparison to the work in \cite{Scheiner2019}, the task of finding a reference track in the radar data is more difficult.
The requirements are stricter in this case because overestimating the area corresponding to the outlines of the VRUs is more critical.
Therefore, this article aims to incorporate the IMU measurements after all.
In particular, it is shown how IMU data can be used to improve the accuracy of separating VRU data from surrounding reflection points and how a fine-tuned version of the internal position filtering is beneficial in rare situations.
The article consists of two major contributions.
First, the proposed system for generating a track reference is introduced.
Second, the GNSS reference system is used to analyze real world VRU behavior.
Therefore, the advantage of measuring stable object centers is used to generate object signatures for pedestrians and cyclists which are not based on erroneous tracking algorithms, but are all centered to a fixed reference point.

\section{Ground Truth Estimation}
\label{sec:system}
\begin{figure}[tb]
\centering
\includegraphics{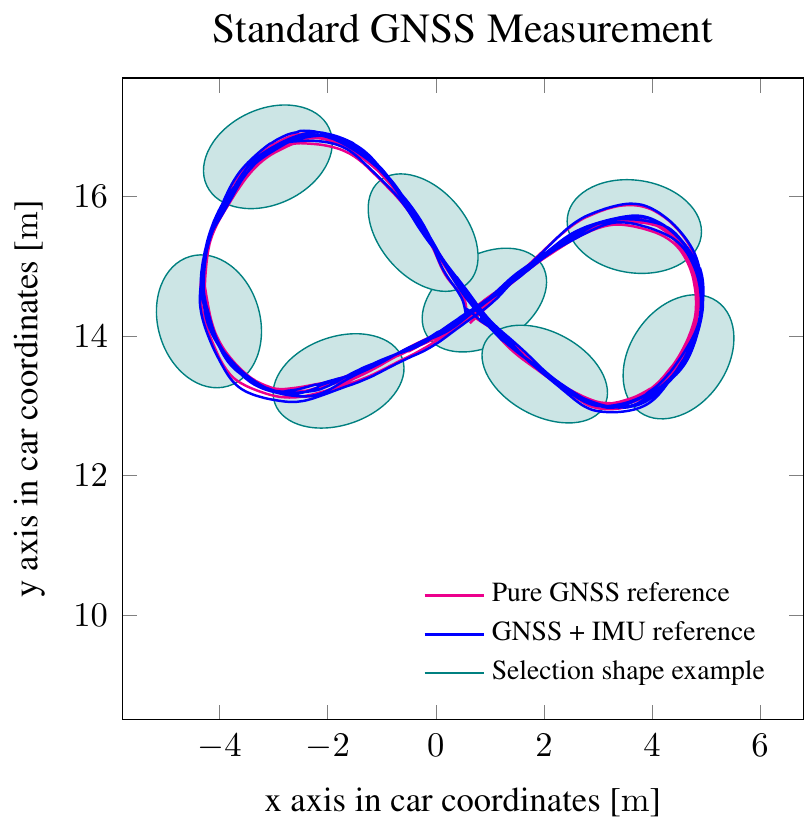}
\includegraphics{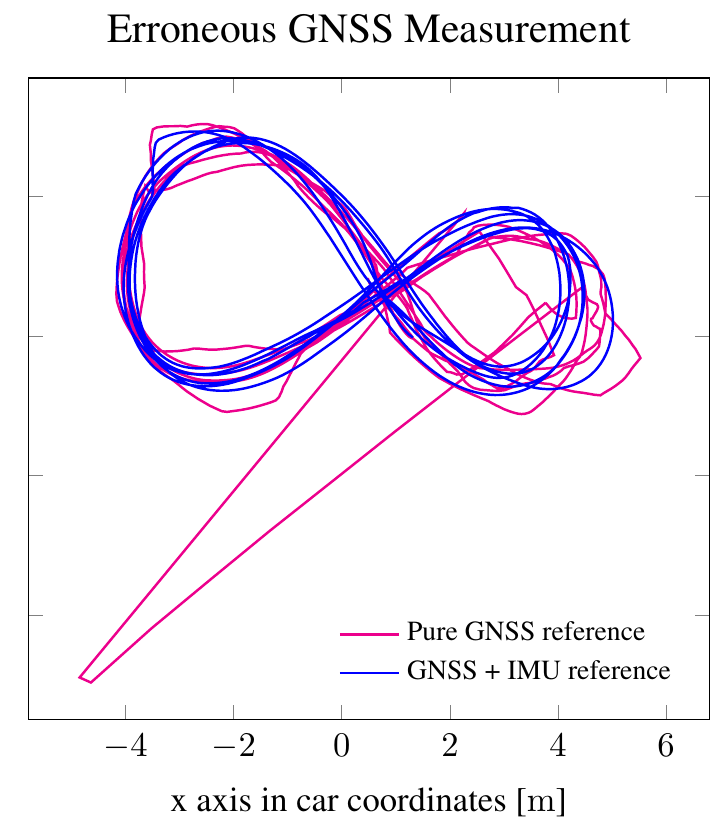}
\caption{Five repetitions of estimated reference system trajectories based on GNSS position with and without usage of IMU. The left plot depicts a normal measurement cycle with smoothed GNSS signal. On the right, the benefits of a well calibrated position filter are visible when taking into account IMU data.}
\label{fig:both_eights}
\end{figure}

According to \cite{Scheiner2019}, the proposed reference system consists of the following components:
VRUs and vehicle are equipped with a device combining GNSS receiver and an IMU for orientation estimation each.
VRUs comprise pedestrians and cyclists for this article.
The communication between car and the VRU's receiver is handled via Wi-Fi.
The GNSS receivers use GPS and GLONASS satellites and real-time kinematic (RTK) positioning to reach centimeter-level accuracy.
RTK is a more accurate version of GNSS processing which uses an additional base station with known location in close distance to the desired position of the so-called rover \cite{Thomas2010}.
It is based on the assumption that most errors measured by the rover are essentially the same at the base station and can, therefore, be eliminated by using a correction signal that is sent from base station to rover.
All system components for the VRU system except the antennas are installed in a backpack including a power supply.
The GNSS antenna is mounted on a hat to ensure best possible satellite reception, the Wi-Fi antenna is attached to the backpack.
Especially for the ego-vehicle, a complete pose estimation (position + orientation) is necessary for the correct annotation of global GNSS positions and radar measurements in sensor coordinates.
For a complete track reference, the orientation of the VRU is also an essential component.
Furthermore, both vehicle and VRU can benefit from a position update via IMU if the GNSS signal is erroneous or simply lost for a short period.
Experiments in \cite{Scheiner2019} revealed that the standard configuration of the internal position filter, which fuses both signals in the GNSS + IMU unit, is not well equipped for unsteady movements of VRUs, especially not for pedestrians.
This quickly led to accumulating positioning errors.
The internal filter uses several heuristics about minimum velocities and turning rates that are required for initialization and standstill detection.
Exemplary trajectories of combined GNSS + IMU positioning with fine-tuned filter parameters versus pure GNSS can be found in Fig. \ref{fig:both_eights}, along with some examples of the data selection area which will be explained in the remainder of this section.
It is clearly visible that combined GNSS + IMU and pure GNSS trajectory both remain on the preset eight-shaped course for regular measurements as depicted in the left plot of Fig. \ref{fig:both_eights}.
In rare cases, the GNSS signal itself contains prominent errors.
These errors may result, e.g., from multipath reflections, satellite occlusion, or high ionospheric activity.
During the measurement campaign for this article this situation occurred only once within roughly ten recorded sequences.
While such measurements would need to be repeated for pure GNSS processing, the second plot in Fig. \ref{fig:both_eights} outlines the benefits of IMU-based position correction for erroneous GNSS measurements.
A quantitative comparison of both methods is given in Sec. \ref{sec:eval}.

\begin{figure}[b]
  \centering
  \includegraphics[width=.89\linewidth]{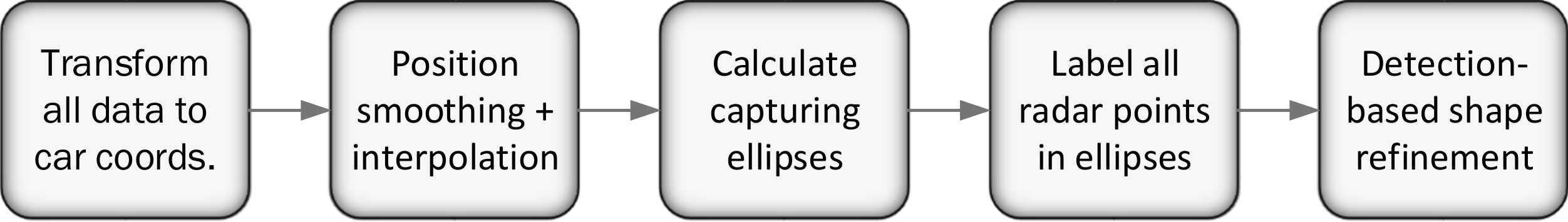}
\caption{Processing steps of ground truth estimation strategy.}
\label{fig:proc_chain}
\end{figure}

Once all data from GNSS, IMU and radar is captured, VRU tracks have to be assigned to corresponding radar reflections.
A basic overview is given in Fig. \ref{fig:proc_chain}.
At first, all data needs to be transformed to a common coordinate system, e.g., car coordinates.
Then, GNSS, IMU, and combined information is smoothed with a moving average filter of length $9$ to remove jitter in positions and movements.
The length corresponds to roughly \SI{0.5}{\second} for GNSS and \SI{0.1}{\second} for IMU data.
Both durations are a trade-off between good smoothing characteristics and the expected interval of continuous VRU behavior.
For each timestamp of each radar measurement the position is estimated by cubic spline interpolation.
In the next step, an area around the position has to be defined. 
If ordinary car tracks were referenced with this method a rigid selection area could be easily defined as car orientation and outer dimensions can usually be estimated very precisely.
For VRUs though, the selection area is more volatile.
Swinging body parts or turning handlebars of cyclists, for example, complicate defining a fixed enclosing structure.
Hence, two different versions of non-rigid surrounding shapes are proposed for the VRUs under consideration.
For a pedestrian, a Gaussian distribution is assumed, thus an ellipse is used with its major axis oriented in movement direction (yaw angle) $\phi$.
Major and minor axis of the ellipse are calculated from fixed minimum pedestrian dimensions (empirically determined: $\SI{1.5}{\meter} \times \SI{1.2}{\meter}$) plus a variable extra length for swinging body parts defined by its velocity $v$ and yaw rate $\dot{\phi}$:
\begin{align}
  \text{ax}_\text{major}&=\begin{cases}
    \SI{1.5}{\meter} + \min(|v| \cdot \SI{1}{\second}, \SI{1}{\meter}), \qquad \quad \qquad &\text{if $v\geq\SI{0.05}{\meter\per\second}$}.\\
    \SI{1.5}{\meter}, \qquad \quad \qquad &\text{otherwise}.
  \end{cases}\\
  \text{ax}_\text{minor}&=\begin{cases}
    \SI{1.2}{\meter} + \min(|\dot{\phi}| \cdot \SI{5}{\meter\second\per\radian}, \SI{1}{\meter}), &\text{if $v\geq\SI{0.05}{\meter\per\second}$}.\\
    \SI{1.5}{\meter}, &\text{otherwise}.
  \end{cases}
\end{align}

$\dot{\phi}$ can be directly adopted from the IMU data.
For a stable position filtering, the reference system is also capable to return good estimates for $\phi$ and $v$ which makes the more complicated derivation from pure GNSS data in \cite{Scheiner2019} obsolete.
The cyclist is labeled inside a rectangle with fixed length of \SI{2.5}{\meter} oriented in driving direction $\phi$ and width of \SI{1.2}{\meter} plus a variable amount based on its yaw rate.
\begin{equation}
     \text{width}_\text{rect} = \SI{1.2}{\meter} + \min(|\dot{\phi}| \cdot \SI{5}{\meter\second\per\radian}, \SI{1}{\meter})
\end{equation}
As bikes usually cannot turn without driving, the derivation of $\phi$ assumes constant continuation of the cyclist's orientation for $v<\SI{0.05}{\meter\per\second}$ overruling the estimated yaw angle of the reference system.
At each time step all radar detections that lie inside the defined regions are being assigned the corresponding track.
Lastly, the enclosing shape is stored as a bounding box that fits a symmetric ellipse-shaped area around the object center which is aligned with the VRU's orientation $\phi$ and includes all radar points of the current time step.
This ensures that overestimating the outlines in previous steps has a lower impact on the estimated radar track.

\section{VRU Object Signatures}
\label{sec:vru}
\begin{figure}[htb!]
\centering
\begin{minipage}{.48\textwidth}
  \centering
  \includegraphics[width=.95\textwidth]{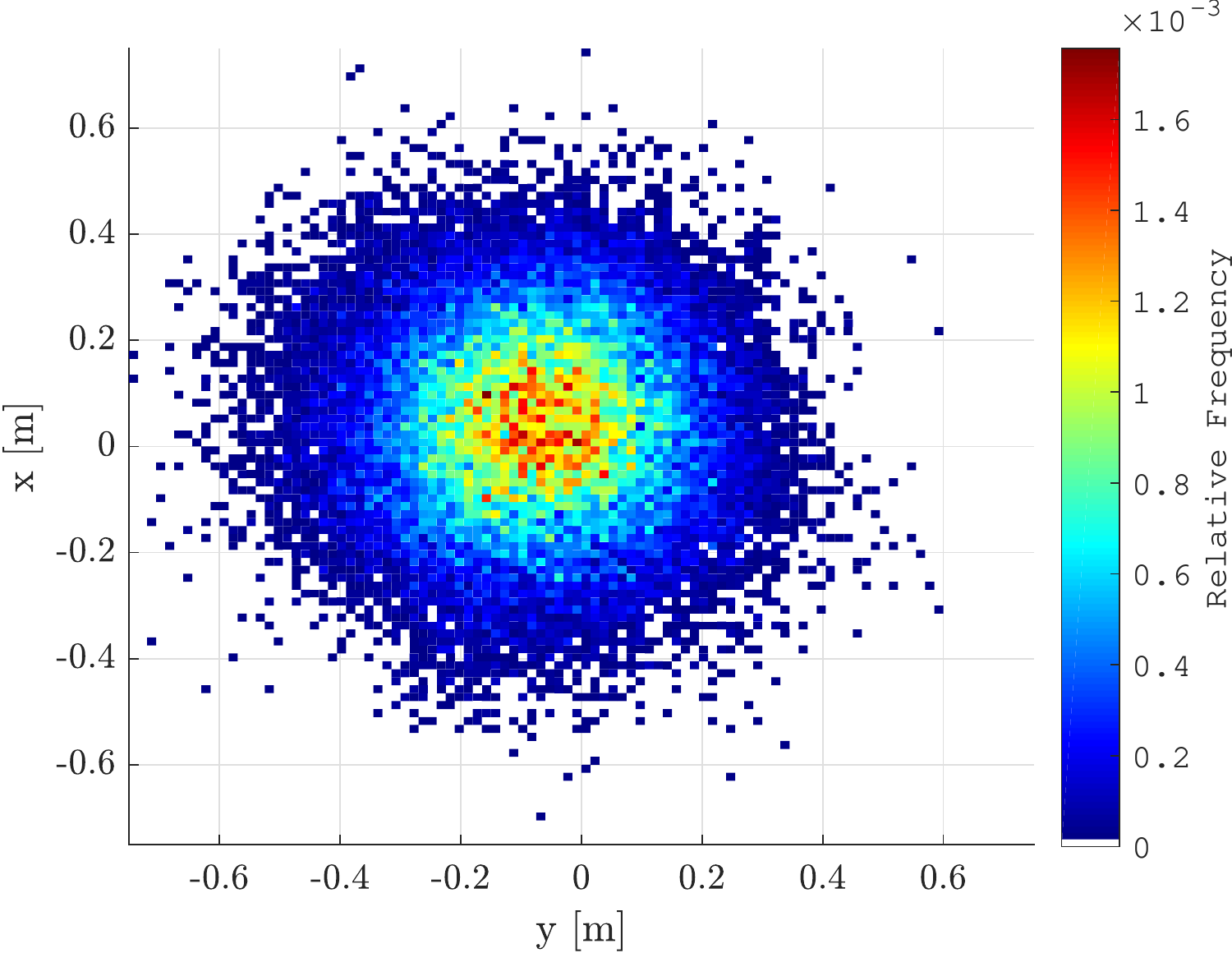}
  \caption{Pedestrian radar object signature calculated relative to GNSS+IMU position measurements.}
  \label{fig:pedSignature}
\end{minipage}
\hfill
\begin{minipage}{.48\textwidth}
  \centering
  \includegraphics[width=.95\textwidth]{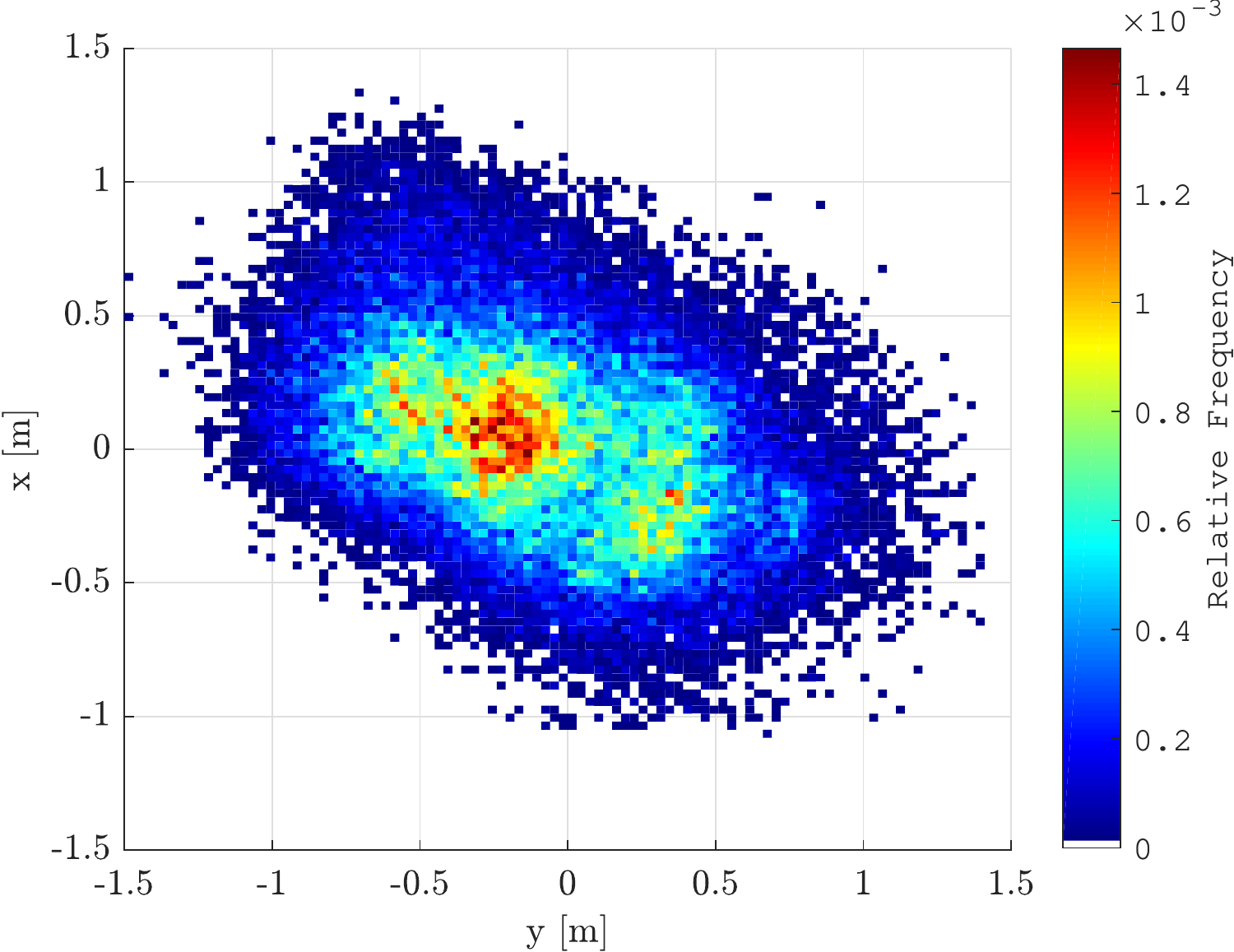}
  \caption{Cyclist radar object signature calculated relative to GNSS+IMU position measurements.}
  \label{fig:cycSignature}
\end{minipage}
\end{figure}

Sophisticated and sensor specialized measurement models allow Extended Object Tracking and fusion to obtain a higher precision and higher robustness as they allow a better interpretation of the measurements.
Also, modeling errors can be significantly reduced.  
The distribution of car measurements can, e.g., be modeled as Gaussian distributed over the cars extent or they can be modeled more sophisticatedly using several reflection centers \cite{Granstrom2017}.
Therefore, obtaining specialized measurement models of VRUs is particularly important.
Knowing the position of an observed object in a traffic scenario allows to extract its detections in every time step and transform them into an observed object coordinate system, where the x-axis corresponds to the direction of movement. 
The target signature of a pedestrian in Fig. \ref{fig:pedSignature} is obtained by accumulating all measurements over time in this coordinate system.
All measurements of an eight shaped movement as depicted in Fig. \ref{fig:both_eights} are evaluated to include all aspect angles. 
In contrast to the target signatures shown in \cite{Haag2018}, the target signature is closer to a round Gaussian distribution, but without a clear peak at the object center. 
Frequent measurements occur in a circular area with \SI{25}{\centi\meter} diameter around the origin.
In contrast to cyclist and car signatures, the major axis is not clearly headed towards the movement direction.
This obstructs the estimation of the object's orientation.
In comparison, the cyclist in Fig. \ref{fig:cycSignature} is more elongated in driving direction.
Both wheels seem to build two different peaks and also pedals and legs are roughly indicated.
The cyclist signature is bent rightwards in its heading direction. 
This might be caused by GNSS inaccuracies or, more likely, one loop provides more radar detections than the other half of the eight due to a position in the scanning area with higher resolution facilitating more detections per object per scan. \\
The presented measurement models have to be further adjusted to sensor specific influences.
The impact of varying aspect angles and distances have to be examined to create non-static sensor specific measurement models.
Furthermore, a sophisticated range rate measurement model has to be developed for pedestrians and cyclists adjusted to non-uniform leg and arm movements.
The GNSS+IMU position measurements can be exploited to obtain those sophisticated measurement models with real world data.  

\section{System Evaluation}\label{sec:eval}
A series of experiments was conducted to evaluate the performance of the proposed reference system.
In order to get measurements from all angles of the VRU, an eight-shaped path was chosen (cf. Fig \ref{fig:both_eights}) and evaluated in three different scenarios consisting of more than 8500 radar cycles using two chirp sequence radar sensors operating at \SI{77}{\giga\hertz} with resolutions of approximately \SI{15}{\centi\meter} in range, \SI{2.4}{\degree} in azimuth angle, and \SI{0.17}{\meter\per\second} in radial velocity:

1) \hspace{0.3cm} Pedestrian walking at constant normal speed (with/without reference)\\
2) \hspace{0.3cm} Cyclist driving at approximately \SI{3}{\meter\per\second} (with/without reference)\\
3) \hspace{0.3cm} Cyclist driving at approximately \SI{3}{\meter\per\second} with GNSS perturbations (with reference)

\begin{figure}[tb]
\centerline{\includegraphics{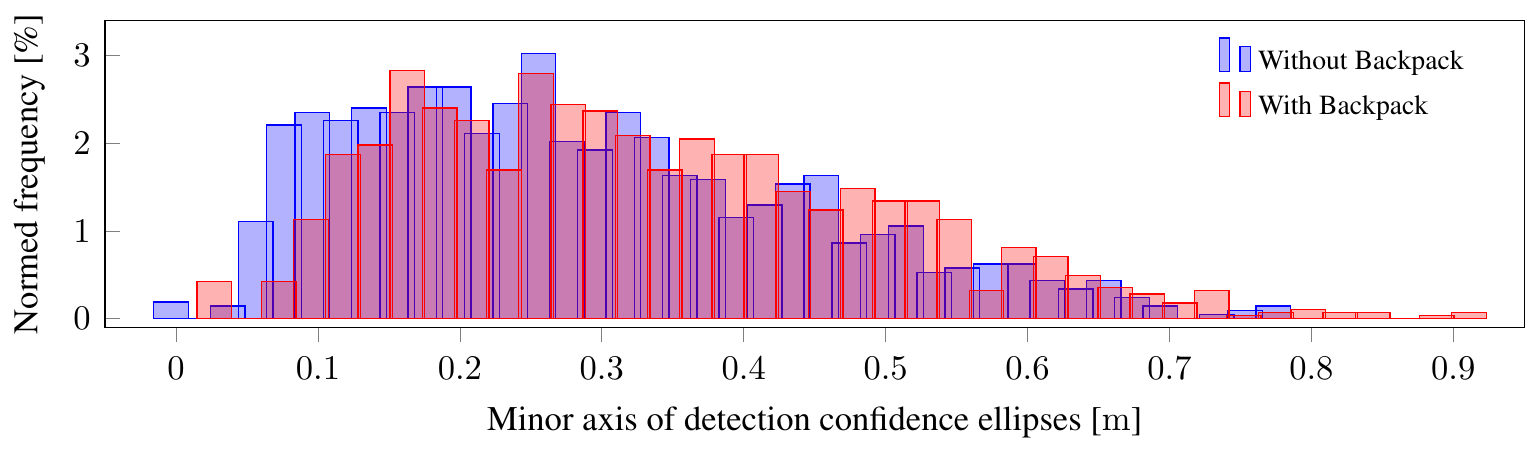}}
\caption{Histogram of a pedestrian's extension along minor axis of confidence ellipse -- scenario 1).}
\label{fig:ped_hist}
\centerline{\includegraphics{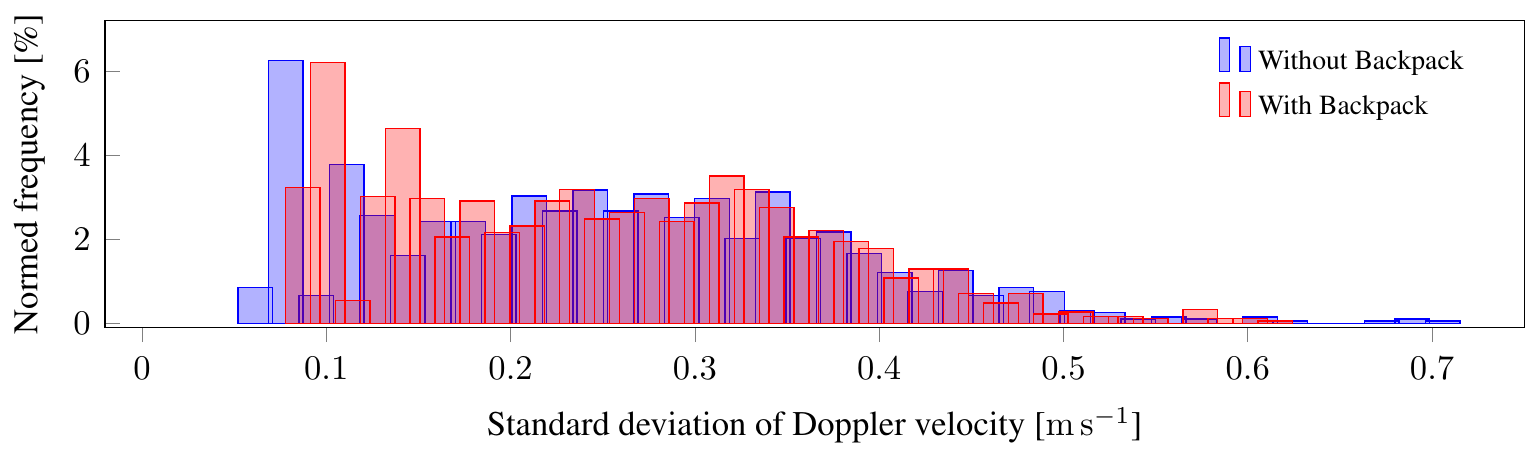}}
\caption{Histogram of a cyclist's standard deviations in mean Doppler velocity -- scenario 2).}
\label{fig:bike_hist}
\end{figure}

Scene 3) occurred only by chance and is, hence, only available with the reference system.
All measurements were hand-labeled by a human expert and additionally all measurements including a GNSS reference were annotated automatically with the proposed method.
Several indicators are important for comparing the proposed method with conventional manual labeling.
First, the accuracy of the method has to be compared against the ground truth obtained from manual labeling.
Second, the differences in measured values for VRU carrying or not carrying the reference system have to be estimated.\\
To determine the performance of the point-to-track assignment system, two scores were calculated.
Let $\TP$ (true positives) be the amount of points correctly assigned to a VRU, $\FP$ (false positives) the incorrectly assigned points, and $\FN$ (false negatives) the amount of points that incorrectly have not been included in a track.
Then, the precision of the method can be calculated as $\mathit{Pr} = \TP/(\TP+\FP) \in [0,1]$ and the recall is $\mathit{Re} = \TP/(\TP+\FN) \in [0,1]$.
The scores are calculated for scenario 1) and 2) with attached GNSS reference.
The macro-averaged results, i.e., averaged individual scores yield a \textbf{precision} of \textbf{\SI{99.53}{\percent}} and a \textbf{recall} of \textbf{\SI{99.61}{\percent}} (cf. precision of \SI{99.48}{\percent} and recall of \SI{99.66}{\percent} in \cite{Scheiner2019}).
Note, that it would certainly be possible to improve these scores for this data set by fine-tuning the parameters of the selection area.
This could, however, easily result in an overfitting on the given data set, i.e., a parameterization that would not generalize well on other data.
Both scores, precision and recall, are identical up to the second decimal place for pure GNSS and combined GNSS + IMU referencing when only regarding the first two scenarios.
In scene 3) the fluctuations of the GNSS signal deteriorate the selection scores to a \textbf{precision} of \SI{97.35}{\percent} and a \textbf{recall} of \SI{86.41}{\percent} when using the pure GNSS trajectory.
However, the combined GNSS + IMU position yields a \textbf{precision} of \SI{99.49}{\percent} and a \textbf{recall} of \SI{97.71}{\percent} which is close to perfect.
Despite generally high accuracies for both methods, these findings suggest a higher stability against adverse environmental conditions when using a combined GNSS + IMU reference system.
This is especially beneficial for a wider series of experiments as it results in a lower rejection rate of measurement files.\\
In order to determine how wearing the GNSS equipment alters measured values, manually labeled data of scenes 1) and 2) are compared for scenarios where the reference system was and was not worn.
Important criteria for comparison are measured amplitudes, variations of Doppler values, the spatial extent, and the amount of detections per measurement.
Therefore, in \cite{Scheiner2019} the mean reflected power compensated for free-space path loss using $R^4$ correction, the standard deviation of Doppler values, the length of the major and minor axis of the \SI{95}{\percent} confidence ellipse, and the amount of detections weighted by the mean distance to the sensor are calculated for each measurement cycle.
Unpaired t-tests on all data in \cite{Scheiner2019} revealed the only statistically significant differences in the length of the minor axis of the \SI{95}{\percent} confidence ellipse of pedestrians and for the standard deviation of Doppler values for cyclists.
In order to closer evaluate the found differences Figs. \ref{fig:ped_hist} and \ref{fig:bike_hist} display the distribution of minor axis lengths and Doppler standard deviations, respectively.
For the pedestrian it can be seen, how the distribution of values for minor axis of the confidence ellipse of all radar detections is slightly shifted towards larger values.
This makes sense as this axis usually corresponds to sagittal body direction which is directly elongated by the reference backpack.
The same effect on the extent of a pedestrian should, however, also occur from any other ordinary backpack with e.g. a laptop inside.
For the cyclist as similar behavior of radial velocity standard deviations can be recognized, i.e., with the exception of the higher regions above \SI{0.5}{\meter\per\second} the backpack distribution is shifted towards bigger values.
This is less intuitive as it would be expected that a backpack enlarges the quasi-static torso and, therefore, leads to smaller Doppler deviations.
A simple explanation for this behavior would be different driving speeds during these scenarios.
In any case, the distributions still look very similar, suggesting that the observed differences introduce a bias, but unlikely make them less relevant.

\section{Extended Object Tracking }
EOT identifies filtering techniques that supplement point tracking by tracking the objects' spatial extents and their orientations over time.
It is sufficient for EOT in traffic scenarios to approximate the road users extents with basic shapes such as ellipses, rectangles or circles.
In Section \ref{sec:vru} and \cite{Haag2018}, it was shown that ellipses are suitable forms for VRUs.
Furthermore, as the accumulated target measurements are Gaussian distributed over the object's extent the Random Matrix Model (RMM) \cite{Koch2008}, \cite{Feldmann2011} seems promising for EOT of VRUs on the provided data.
The RMM complements the centroid and kinematic state vector $\textbf{x}_k$ with a symmetric positive definite matrix $X_k$ that represents the object's extent.
The object's length, width and orientation is obtained by principal component analysis of the extent matrix.
The given scenario is very challenging due to the non-linear movement.
Therefore, the RMM is combined with the constant turn motion model \cite{Zhai2014} and adjustments of the extent matrix according to the object motion \cite{Lan2012a}.

\begin{figure}[h!]
\centering
\includegraphics[width=.95\textwidth]{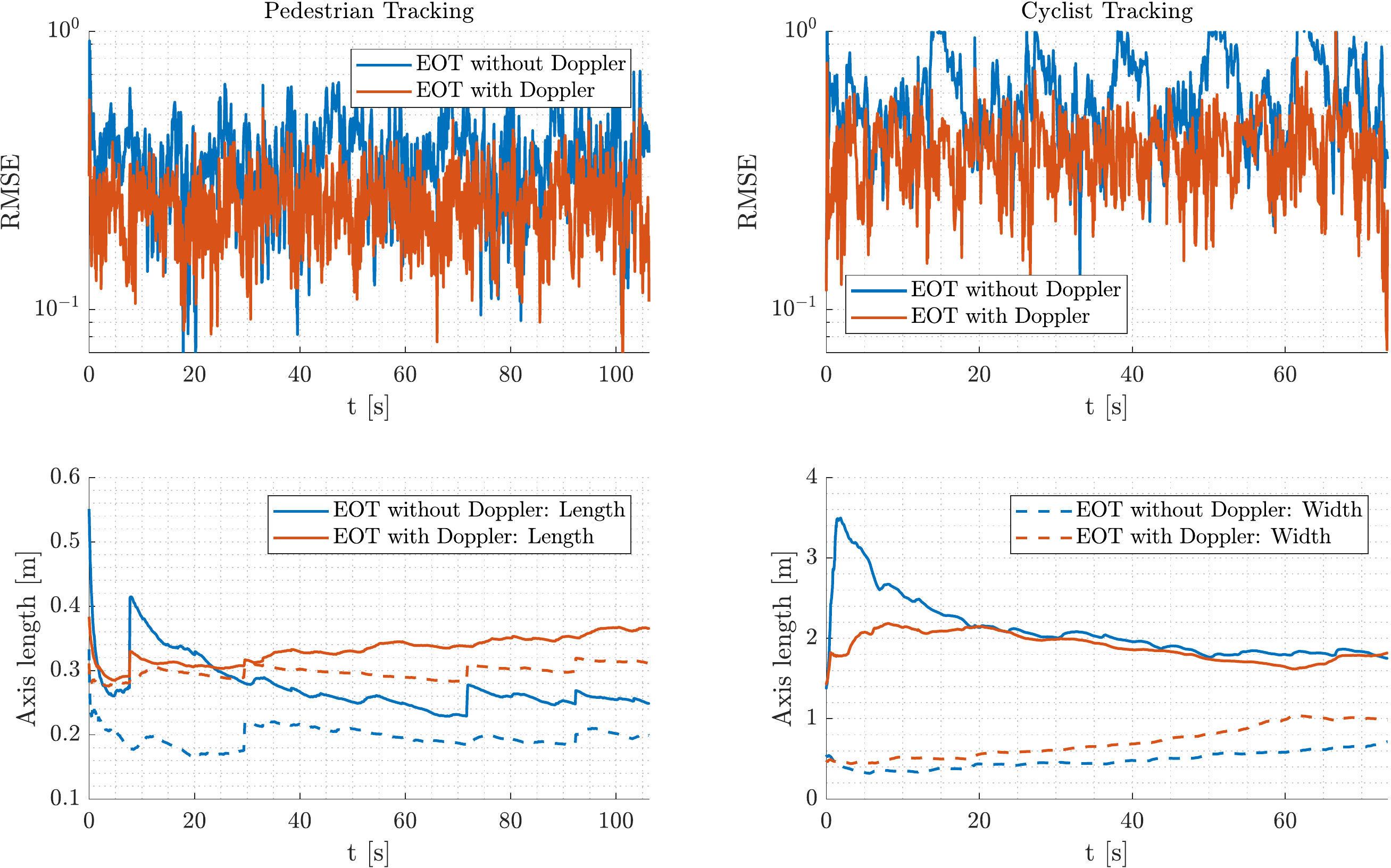}
\caption{RMM results: RMSE of centroid, length, and width of a pedestrian and a cyclist on an eight shaped path.}
\label{fig:EOTerr1}
\vspace{0.7cm}
\centering
\includegraphics[width=.95\textwidth]{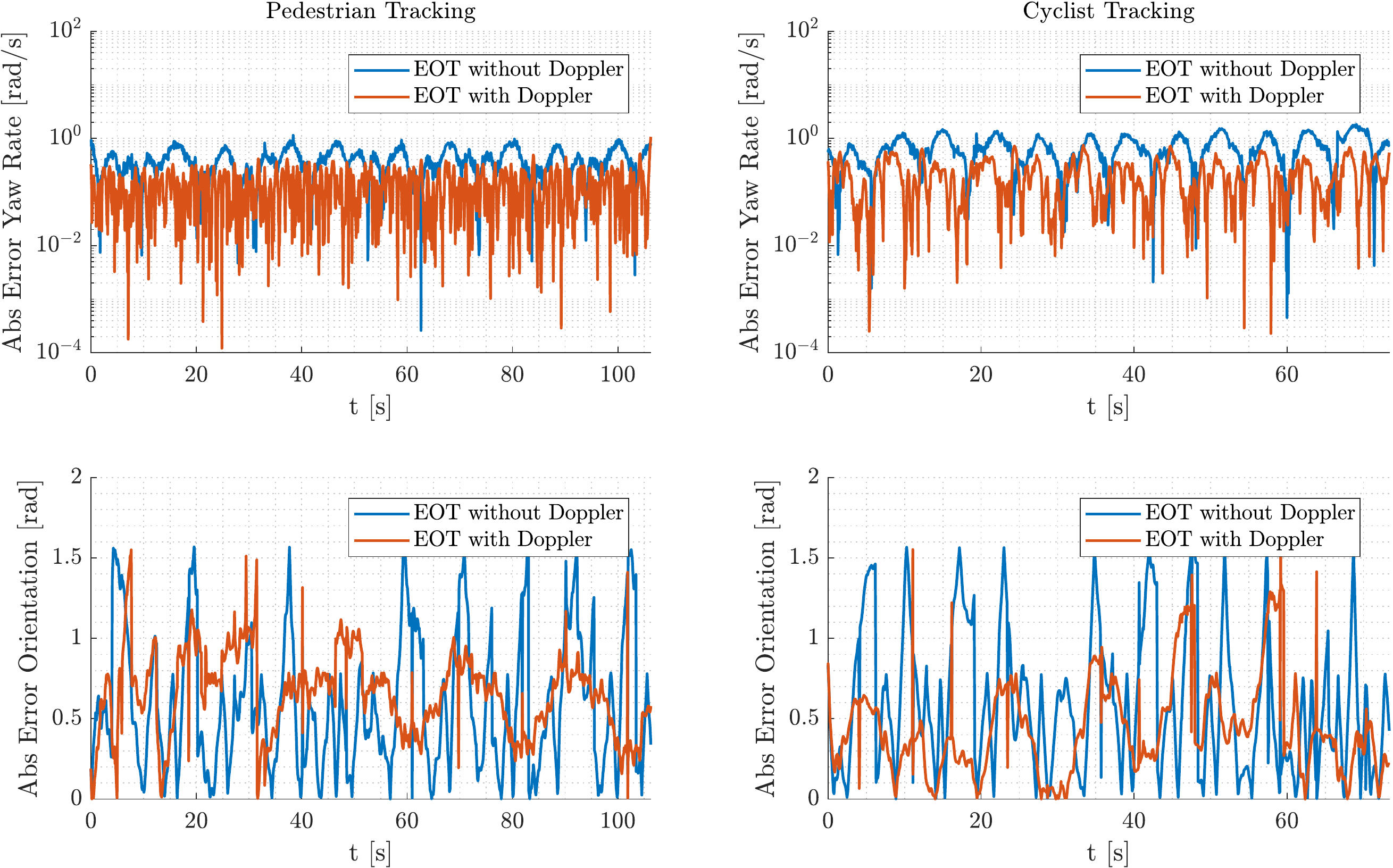}
\caption{RMM results: Absolute yaw rate error and absolute orientation error of a pedestrian and a cyclist on an eight shaped path.}
\label{fig:EOTerr2}
\end{figure}

Fig. \ref{fig:EOTerr1} shows the results of pedestrian and cyclist tracking on the eight shaped walk/ride scenarios 1) and 2).
Both VRU centroids can be tracked accurately.
Utilizing the Doppler velocity measurements reduces the RMSE drastically.
Length and width are well estimated when the Doppler measurements are incorporated. 
Without them, stability is lost in the first \SI{20}{\second}.
It can be observed that wrong prior orientation causes changes in axis length first instead of a rotation.
The tracking framework is optimized to maintain stable yaw rate and, thereby, orientation estimation.
This allows to handle non-linear movements and reduce orientation errors in contrast to \cite{Haag2018} where non-linear movements caused estimation errors.
Fig. \ref{fig:EOTerr2} depicts the absolute yaw rate and orientation errors.
Utilizing Doppler values the yaw rate is determined very accurate but with time delay.
The orientation errors are very high for both VRUs.
This shows that estimating the yaw rate accurately is not sufficient for the determination of object orientation.
Hence, specific measurement models are needed to gain a stable orientation in challenging scenarios.

\newpage

\newpage
\section{Conclusion}\label{sec:conclusion}
In this article, a method for fast reference generation for automotive radar tracking of VRUs was proposed.
The system is based on the combination of two GNSS receivers mounted on the ego-vehicle and the VRU in combination with an IMU.
Radar data is automatically assigned to a track if it falls within a close area around the VRU's estimated position.
The selection area is determined by the kind of tracked VRU, i.e., pedestrian or cyclist, its speed and yaw rate.
Experiments prove the accuracy of the proposed method with precision and recall both over \SI{99}{\percent}.
The supplementary IMU upheld the scores are over \SI{98}{\percent} during perturbations in the GNSS signal.
The article concludes by using the generated reference to evaluate Extended Object Tracking on VRUs.
By collecting more data sophisticated measurement models can easily be generated.
This allows developing more precise, robust, and faster extended object tracking methods in order to provide a higher safety level for all road users.
Besides creating better VRU models, it is also planned to use several GNSS backpacks for tracking multiple VRUs in future work.
This involves adapting the selection strategy to cope with situations where selection areas overlap.

\vspace{1em}
\bibliographystyle{IEEEtran}
\bibliography{IEEEabrv,mybibfile,library}

\end{document}